\documentclass{article}
\usepackage{amsmath}
\usepackage{booktabs, tabularx}
\usepackage[tableposition=top, font=bf]{caption}
\usepackage{booktabs,natbib}
\usepackage{graphicx}
\usepackage{libertinus}
\usepackage{microtype}
\usepackage[hidelinks]{hyperref}
\title{The Popularity-Homophily Index: A new way to measure Homophily in
Directed Graphs}

\author{Shiva Oswal\\\normalsize
UC Berkeley Extension,
Berkeley, CA, USA}

\date{}

\begin{document}
\maketitle

\begin{abstract}
\noindent
In networks, the well-documented tendency for people with similar
characteristics to form connections is known as the principle of
homophily. Being able to quantify homophily into a number has a
significant real-world impact, ranging from government fund-allocation
to finetuning the parameters in a sociological model. This paper
introduces the ``Popularity-Homophily Index'' (PH Index) as a new metric
to measure homophily in directed graphs. The PH Index takes into account
the ``popularity'' of each actor in the interaction network, with
popularity precalculated with an importance algorithm like Google
PageRank. The PH Index improves upon other existing measures by
weighting a homophilic edge leading to an important actor over a
homophilic edge leading to a less important actor. The PH Index can be
calculated not only for a single node in a directed graph but also for
the entire graph. This paper calculates the PH Index of two sample
graphs, and concludes with an overview of the strengths and weaknesses
of the PH Index, and its potential applications in the real world.
\end{abstract}


\section{Introduction}

Homophily, meaning ``love of the same,'' captures the tendency of
agents, most commonly people, to connect with other agents that share
sociodemographic, behavioral, or intrapersonal characteristics
(McPherson et al, 2001). In the language of social interaction networks,
homophily is the ``tendency for friendships and many other interpersonal
relationships to occur between similar people'' (Thelwall, 2009). The
social, economic, and political outcomes of homophily are significant.
In urban settings for instance, homophilic tendencies in people tend to
create exclaves of a highly concentrated race or ethnicity (Xu et al,
2021). Chinatown in San Francisco and Little Italy in New York, are
prime examples of homophily in action. Race, socioeconomic factors, and
religion all play a prime role in determining the likelihood of two
people communicating with one another, or living nearby each other (Xu
et al, 2021).

Identifying and measuring homophily can have significant real-world
consequences. Identifying homophily in housing can help authorities deal
with dogged pockets of segregation, while identifying homophily in
social media can reveal the importance of certain attributes and users,
aiding social media companies, advertisers, and others (Weng et al,
2010). Measuring homophily provides numeric detail to the above
situations, which in turn can influence the amount of funding allocated,
and resources spent in response. For these reasons, an effective measure
of homophily is needed. The most nuanced models need to take the
structure of the graph into consideration.

This paper presents a new way of measuring the homophily of an attribute
in a directed graph --- the ``Popularity-Homophily Index'' or PH Index
for short. The paper then looks at two example datasets, calculates the
PH Index in both cases, and then compares their results.

The first dataset is a social network of Github developers. The second
dataset is an internet network of links between political blogs in the
lead-up to the 2004 US presidential election. These two graphs are
radically different in their contents, but this paper finds that the PH
Index has relevance in both. For sufficiently complex directed graphs,
the PH Index is significantly more powerful than other related measures
of homophily in common use, such as the classic External-Internal Index
(EI Index).

\section{Related Work}

The idea and motivation to measure homophily in directed graphs is not
new. For example, Twitter's social network can be modelled as a directed
graph, with users as nodes, and follower relations as directed edges.
Measuring homophily was an important part of the development of
TwitterRank, which measured the influence of prominent Twitterers in a
sample dataset (Weng et al, 2010). Other papers have sought to determine
the extent of the political ``echo chamber'' on social media such as
Twitter. After classifying users as either Democratic or Republican
based on the content of their tweets, one paper then sought to measure
the political homophily present in the resulting graph (Colleoni et al,
2014). This measurement allowed them to quantify the scale and impact of
the political echo chamber on social media. Additionally, ``Degree
Weighted Homophily'' (DWH) takes the structure of the graph into
account, and provably gives a lower-bound for the convergence time of
certain simulations, where nodes represent ``agents'' that are either
attracted to or repelled by their neighbors (Golub and Jackson, 2012).
These simulations are often excellent models for the behavior of
real-world people, such as the tendency of people of the same race to
cluster in the same neighborhoods. However, DWH is nearly impossible to
compute efficiently for real-world graphs, since its basic formulation
takes exponential time to compute.

A recently developed and widely used measure of homophily is called the
\emph{Assortativity Coefficient} \(r\), which satisfies the formula
\(r = \frac{\mathrm{tr}(E) - \mathrm{sum}(E^{2})}{1 - \mathrm{sum}(E^{2})}\) (Newman, 2003).
Easy and efficient to calculate, the assortativity coefficient has been
widely used in research into homophily (Chang et al, 2007). In the case
of a binary classification, \(E\) is a 2x2 matrix containing the
fraction of each possible type of edge in the graph. For example, if
every node has a color attribute that is either ``black'' or ``white'',
the fraction of edges that start at a ``black'' node and end at
``white'' node is one of the 4 possible kinds of edges. Additionally,
$tr$ is the trace of a matrix (the sum of its main diagonal), and
$sum$ is the sum of the elements of the entire matrix.

A significant issue with the assortativity coefficient is that it does
not take the overall structure of the graph into account since it
reduces the entire graph into a matrix, losing much of the intricate
complexity of the graph.

All of these measures of homophily have their advantages and drawbacks.
The PH Index is influenced by many of these measures, and is designed to
be versatile, easy, and efficient to calculate, while nuanced enough to
take the entire structure of the graph into account.

\section{Background}

Define an edge to be a \emph{homophilic edge} if it connects two nodes
that share some attribute, and define it be a \emph{non-homophilic edge}
if otherwise. For the purposes of this paper, assume that every node
attribute is binary, and can be represented as either 0 or 1. Many
non-binary attributes can be converted into binary attributes with a bit
of extra work. For example, if every node is a person with their own
``height'' attribute, we can create a binary attribute of `tallness'.
Arbitrarily define a node to be `tall' if its height attribute is more
than 6 feet, and `not tall' if otherwise.

Define a pair of \emph{similar nodes} to be two nodes (not necessarily
connected by an edge) that have the same value for a given attribute,
either both 0 or both 1. For a given node $n$, one of the oldest,
simplest, and most powerful measures of homophily is called the EI Index
or alternatively the Coleman Homophily Index since it was introduced in
his landmark book \emph{Introduction to Mathematical Sociology}
(Coleman, 1964).

In the set of all node \emph{n}'s out edges (in the case of a directed
graph), let \(e_{n}\) represent the number of ``external''
non-homophilic edges, and \(i_{n}\) represent the number of ``internal''
homophilic edges. Then the \emph{EI Index} is defined by Equation \ref{ei_index}.

\begin{equation}
\label{ei_index}
\mathrm{EI} = \frac{e_{n} - i_{n}}{e_{n} + i_{n}}
\end{equation}

%Equation : Definition of the EI Index

Note that the EI Index is -1 for a completely homophilic node (all of
its edges are homophilic edges), and 1 for a completely heterophilic
node (Coleman, 1964). The EI Index for the entire graph is simply the
average of the EI Index for each individual node. Most random graphs
that do not demonstrate homophily tend to have a value of the EI Index
of approximately 0.

For the purposes of the PH Index, using the \emph{Weighted EI Index}
(WEI Index) is more appropriate. The necessity of the WEI Index stems
from the fact that in a binary classification, the total number of nodes
with one value of an attribute may differ significantly from the total
number of nodes with the other value of that same attribute. For
example, consider a social network of 200 individuals, with 180 men and
20 women. If Bob (a male) and Jennifer (a female) both are connected to
a single man and a single woman, then their EI indices will both equal
\(0.0\). However, common sense suggests that Jennifer exhibits a greater
amount of homophily, since there are very few other women she could be
connected to.

Influenced by the above observation, define the weight \(w\) of node
\emph{n} to be the ratio of the number of non-similar nodes in the
entire graph, divided by the number of similar nodes. Therefore, Bob's
weight would equal 0.11, and Jennifer's weight would equal 9.0. For node
\emph{n}¸ consider the same variables as in the definition of the EI
Index. The equation for the WEI Index is stated in Equation \ref{wei_index}.

\begin{equation}
\label{wei_index}
\mathrm{WEI} = \frac{\frac{e_{n}}{w} - i_{n}}{\frac{e_{n}}{w} + i_{n}}
\end{equation}

%Equation : Definition of the WEI Index

The Weighted EI Index essentially normalizes the EI Index, calculating
it as if there were an equal number of nodes with each possible value of
the attribute. In our example above, Bob's WEI is \(0.8\) and Jennifer's
WEI is \(- 0.8\). As predicted, after weighting, Jennifer exhibits
strong homophily.

\section{Popularity}

Define the \emph{popularity} of a node to informally represent its
importance or centrality in the graph. In the Twitter social network for
example, both Barack Obama and Kim Kardashian would have a high
`popularity'. There are multiple well-known methods to measure
popularity in a directed graph. A few of the most common types are
listed below.

\subsection{In-Degree Centrality}

In-Degree Centrality is the simplest measure of popularity. It is also
the fastest, with a runtime of \(\mathcal{O}(V + E)\) (Zhang and Luo,
2017). Logically, it follows that a node with many incoming edges is
important. Thus, a node's popularity is simply its in-degree, normalized
by the total number of nodes in the graph.

\subsection{Betweenness Centrality}

Betweenness Centrality defines a node's popularity to be the fraction of
shortest paths that pass through the node. Logically, it follows that
the more central a node is in the graph, the higher the betweenness
centrality, and the higher the importance of that node. The major
drawback of this algorithm is its slowness: without optimization it runs
in \(\mathcal{O}\left(VE\right)\) (Zhang and Luo, 2017).

\subsection{PageRank}

The PageRank algorithm is a direct upgrade on the In-Degree Centrality
algorithm. It has a time complexity of
\(\mathcal{O}\left( k(V + E) \right)\), where $k$ is the maximum
number of iterations (the default in NetworkX is 100) to run the
algorithm for. The intuition behind the algorithm is that since not all
nodes are equally popular, they should not all be weighted the same.
Therefore, not all incoming edges are equal, and they should not be
counted equally. The PageRank algorithm was originally developed to
order web pages for the search engine Google (Page et al, 1999). The
full details of the PageRank algorithm are beyond the scope of this
paper, although a short overview is provided. By treating page-rank as a
``fluid'', at every iteration of the algorithm, each node equally
distributes all of its ``fluid'' to its neighbors. Generally, a node
with many incoming edges tends to accumulate a lot of page-rank,
although each incoming edge provides a different amount of ``fluid''
based on its node's current amount of page-rank (Rogers, 2002). While
the argument to justify PageRank may appear circular, the algorithm
provably converges in most cases. Once convergence is reached, the
amount of page-rank a node has is equal to its popularity.

\section{Popularity-Homophily Index (PH Index)}

The Popularity-Homophily Index, presented and developed in this paper,
is a new method for measuring homophily in a directed graph. The PH
Index for a node is closely related to the EI Index, and more
specifically to the WEI Index. The major difference is that the PH index
takes into account the popularity of a node's set of neighbors.

The first step of the algorithm is to calculate the popularity of every
node using one of the popularity algorithms above and store it in the
array \(P\). Additionally, normalize the array so that
\(\sum_{}^{}P = 1\).

The second step is to redefine the variables \(e_{n}\) and \(i_{n}\)
that appeared in the formula for the WEI Index, as shown in Equation \ref{count}.
Essentially, Equation \ref{count} weights these variables by the popularity of the
edges they accumulate over.

\begin{equation}
  \label{count}
  \begin{aligned}
    e_{n} &= \sum_{i \in \text{External\ Nodes}}^{}P_{i}, \\
    i_{n} &= \sum_{i \in \text{Internal\ Nodes}}^{}P_{i}
  \end{aligned}
\end{equation}

%Equation : Formulae for computing the external and internal count for a node, weighted by popularity

Once \(e_{n}\) and \(i_{n}\) have been re-calculated, the formula for
the PH Index is the same as the formula for the WEI Index.

\emph{Why does taking popularity into account create a more robust index
of homophily?} The reason is, in most real-world graphs, ties to
influential nodes are a greater indication of homophilic tendencies than
ties to less influential nodes. That is, not all out-edges are equally
important. For example, research has shown that people's daily lives are
influenced by the YouTube vloggers that they follow (Ladhari et al,
2020). Additionally, the vlogger's popularity tends to be proportional
to the amount of influence that a vlogger has over a person, meaning
that the most popular vloggers wield an enormous amount of influence
over legions of devoted viewers (Ladhari et al, 2020). Naturally, this
phenomenon extends beyond just vloggers, and it is not a stretch to say
that an object or person's ``popularity'' is fundamental to its role in
society.

Finally, the PH Index can easily be extended to measure the homophily of
an entire directed graph. First compute the PH Index for every node in
\(\mathcal{O}(P + V + E)\) time. \(\mathcal{O}(P)\) represents the time
needed to compute the popularity of every node, and it varies based on
the algorithm used. \(\mathcal{O}(V + E)\) represents the time needed to
actually compute the index for each node. Let \(\mathrm{PH}_{n}\)
represent the value of the PH index for node $n$. The PH Index for
the entire graph is simply a weighted average of the PH Index for every
node. Once again, we need to weight by popularity, since highly popular
nodes have a greater influence on the overall homophily of the graph
than less popular nodes. Since we normalized the popularity array, the
formula is simply the one stated in Equation \ref{pop_index}.

\begin{equation}
\label{pop_index}
\mathrm{PH} = \sum_{n \in \text{Nodes}}^{}{P_{n} \times \mathrm{PH}_{n}}
\end{equation}

%Equation : Definition of the Popularity-Homophily Index

Like the classic EI Index, a PH Index value of -1 implies complete
homophily, while a value of 1 implies complete heterophily. Most
non-homophilic graphs will have a PH index of approximately 0.

\section{Experimental Data}

This paper demonstrates the use of the PH Index on two sample directed
graphs. The first graph is a network of software engineers on Github, a
popular website for storing and sharing code (Rozemberczki et al, 2021).
Two engineers are connected by a directed edge if one ``follows'' the
other. Additionally, as part of the data, each engineer job has a binary
classification and is either a web (0) or a machine learning (1)
developer. This information was scraped from the developer profiles by a
machine learning algorithm (Rozemberczki et al, 2021). The links in the
graph themselves were scraped from the publicly available Github API.
The data was collected by researchers at the University of Edinburgh led
by Benedek Rozemberczki and accessed from the Stanford Network Analysis
Project (Leskovec and Krevl, 2014).

The second graph is a network of political blogs from the run up to the
2004 US Presidential election (Adamic and Glance, 2005). In this graph,
a directed edge represents a hyperlink from one network to the other.
Each blog has a binary classification as either left leaning (0) or
right leaning (1), and the creators of the second graph used a
combination of automated and manual labelling to create each label. The
data was collected by Lada Adamic and Natalie Glance, and accessed from
the KONECT Network Collection (Kunegis, 2013).

For purposes of comparison, two additional attributes were added. For
the Github graph, each developer's username was added. The ``name
length'' attribute was engineered from this information based on whether
or not the developer's username was shorter than 7 characters in length,
an arbitrary number. For the political blogs graph, the second attribute
was purely random, set to either 0 or 1 with equal probability.

Basic information about the two graphs is listed in Table~\ref{basic}. For both
graphs, both attributes are binary variables. Note that assortativity,
the classic measure of homophily, ranges from 1 (complete homophily),
down to 0 (no homophily), which is a different scale than the PH Index.

\begin{table}
\caption{Basic information about Github and Blogs
   graphs}
\label{basic}
\begin{tabularx}{\textwidth}{lXXXXXX}
\toprule
Graph &
Nodes ($V$)
& 
Edges ($E$)
 & 
Attri\-bute~1
 & 
Assorta\-tivity
 & 
Attri\-bute~2
 & 
Assorta\-tivity
 \\
\midrule
Github & 37,700 & 2,89,003 & Developer Type & 0.378 & Name Length
& 0.012 \\\addlinespace
Blogs & 1,224 & 19,025 & Political Lean & 0.823 & Pure Random &
0.001 \\
\bottomrule
\end{tabularx}
\end{table}

\begin{figure}
  \centering
\includegraphics[width=.6\textwidth]{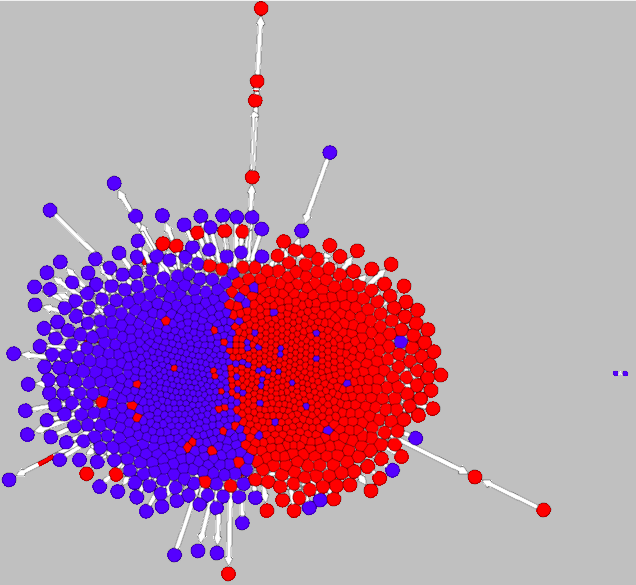}
\caption{Political Lean in the Blog Graph---Blue represents left
leaning blogs, and Red represents right leaning blogs}
\label{pol_lean}
\end{figure}

\begin{figure}
  \centering
	\includegraphics[width=.6\textwidth]{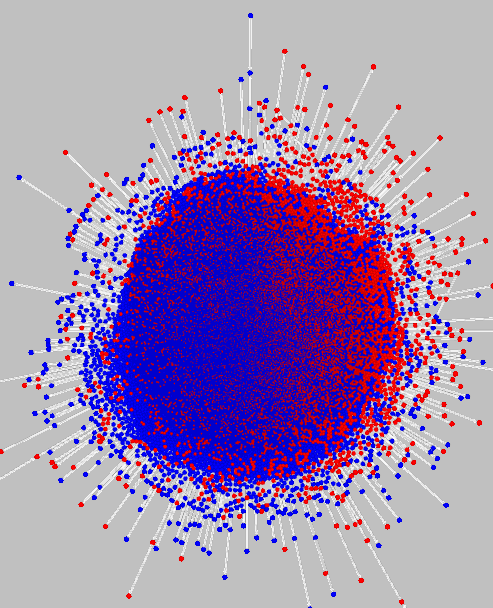}
	\caption{Job in the Github Graph---Blue represents web developers, and
		Red represents machine learning developers}
	\label{job}
\end{figure}

A visualization of both graphs is presented in Figures \ref{pol_lean} and \ref{job}. Note how
homophily is clearly visible in both graphs, although in the Github
graph, the large number of nodes combined with the relatively small
amount of machine learning developers, make it harder to see. The
two-dimensional visualizations were made with Graphia (Freeman et al,
2020).

\begin{figure}
  \centering
	\includegraphics[width=.8\textwidth]{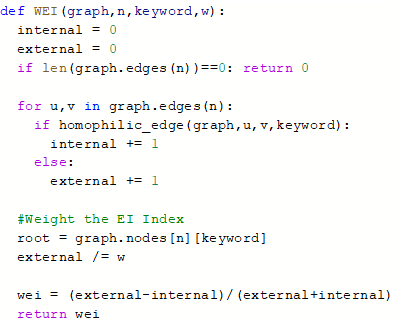}
	\caption{Find the weighted EI Index of node "n" in digraph "graph",
		based on attribute "keyword", with a precomputed weight "w"}
	\label{python_weighted}
\end{figure}

\begin{figure}
  \centering
	\includegraphics[width=.8\textwidth]{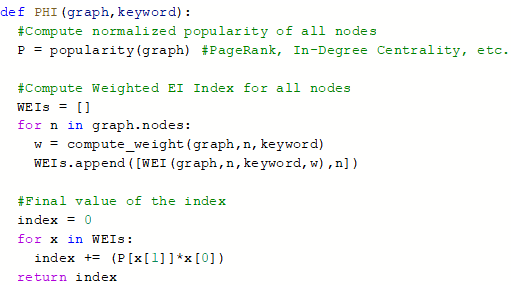}
	\caption{Find the Popularity Homophily Index for DiGraph "graph" and
		attribute "keyword"}
	\label{python_overall}
\end{figure}

To store and manipulate the Github and Blog data, the python module
NetworkX (Hagberg et al, 2008) was used. Both graphs were stored as
NetworkX DiGraphs. All three popularity algorithms mentioned above are
implemented into NetworkX, shortening the code required to compute the
PH Index significantly. Figure~\ref{python_weighted} contains the basic python source-code
for computing the weighted EI Index for a node. Figure~\ref{python_overall} contains the
basic python source-code for computing the overall PH Index for a
directed graph.

\section{Results}

Using the code for the PH index, the homophily of each attribute in each
graph could be measured. The results are organized in Tables~\ref{github} and~\ref{blog}.

\begin{table}
	\caption{Github Graph PH Index}
	\label{github}
	\begin{tabularx}{\textwidth}{l*{3}{X<{\raggedright}}}
		\toprule
		Popularity Type &
		Attribute 1 (Web or ML Developer)
		& 
		Attribute 2 (Short or Long Name)
		& 
		Python Computation Time (seconds)
		\\
		\midrule
		In-Degree Centrality & -0.554 & -0.067 & 0.14 \\
		Betweenness Centrality & -0.537 & -0.003 & 5077 \\
		PageRank & -0.461 & -0.113 & 5.22 \\
		\bottomrule
	\end{tabularx}
\end{table}

\begin{table}
	\caption{Blog Graph PH Index}
	\label{blog}
	\begin{tabularx}{\textwidth}{l*{3}{X<{\raggedright}}}
		\toprule
		Popularity Type &
		Attribute 1 (Left or Right Lean)
		&
		Attribute 2 (Random Number Generator)
		& 
		Python Computation Time (seconds)
		\\
		\midrule
		In-Degree Centrality & -0.738 & -0.030 & 0.14 \\
		Betweenness Centrality & -0.813 & -0.001 & 8.11 \\
		PageRank & -0.709 & -0.031 & 0.84 \\
		\bottomrule
	\end{tabularx}
\end{table}

While weighting by popularity has a huge impact on the final answer,
this impact can be invisible if there are many popular nodes that are
both highly homophilic and highly heterophilic.

As an example, the popularity, calculated with PageRank of every node in
the Github dataset is plotted below in Figure~\ref{page_rank}, using the python
package Matplotlib (Hunter, 2007). Note that there are a handful of
nodes with a huge influence. Consider the most popular 150 developers,
74 of them have a WEI Index below -0.461 (more homophilic than the PH
Index), and 76 of them have a WEI Index above -0.461 (less homophilic
than the PH Index). Naturally, since this split is so even, weighting by
popularity does not have a huge impact on the final answer.

However, we notice a much larger impact of popularity weighting when we
consider the second attribute of the Github graph. A developer is
defined to have a ``short'' username if his username is under 7 letters
long. Approximately 17\% of all developers have a ``short'' username.
However, for some unknown reason, the most popular developers seem to
prefer short usernames, meaning that this ``random'' attribute can and
may actually hold significance.

Out of the top 150 developers, who together account for merely 0.4\% of
the nodes but over 20\% of the total popularity of the graph, an
astounding 28\% have a short username, a statistically significant
difference. Due to the intrinsic nature of PageRank, popular nodes tend
to have edges leading to other popular nodes, meaning that the most
popular nodes tend to be slightly homophilic relative to name length.
For this reason, the weighted PH Index relative to name length is -0.113
as opposed to roughly 0, which clearly indicates some amount of
homophily.

\begin{figure}
  \centering
	\includegraphics[width=\textwidth]{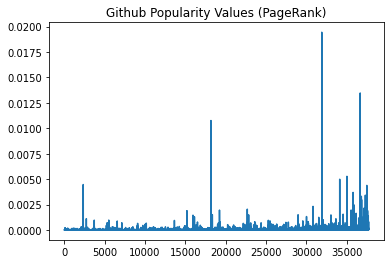}
	\caption{Github Graph Popularity (PageRank)}
	\label{page_rank}
\end{figure}

\section{Conclusion}

The primary purpose of this paper is to introduce the
Popularity-Homophily (PH) Index as a method to compute homophily in
directed graphs. The index was tested on two real-world graphs, the
Github developer dataset, and the Political Blogs dataset. In both
cases, homophily was detected, and as expected from the assortativity
coefficient, the magnitude of homophily in the former graph was smaller
than the magnitude of homophily in the latter graph.

The major advantage of the PH Index occurs in graphs that are dominated
by a limited number of main actors. These actors exert disproportional
amounts of influence on the state of the graph, and in the real-world, a
handful of homophilic connections to popular nodes can easily outweigh a
multitude of heterophilic connections to less popular nodes. The PH
Index has some versatility in that one of many possible algorithms can
be employed in the first step, to calculate the popularity of each node.
In order to decide which algorithm to use, the nature of the real-world
data being analyzed must be considered. For most social networks like
Twitter or Facebook, the PageRank algorithm is probably the most
effective. However, for geographic data, Betweenness Centrality may be
more preferable due to the literal connection between that algorithm and
geography.

As a warning, the magnitude of the PH Index may be a red herring. The
scale is not really linear, as a PH Index of -0.7 demonstrates much
stronger homophily over -0.4 than what might be expected. Additionally,
the magnitude of the PH Index may have a slightly different meaning in
one graph when compared to another graph, since both graphs will have
their own unique structure. For this reason, the main conclusions one
should draw from the PH Index are in terms of relative magnitude.

\section{Future Work}

The full relevance and applications of the novel PH Index remain to be
seen. Below are a few real-world areas of research where the PH Index
may be applicable.

\begin{itemize}
\item
  On social media, are people with one political lean more homophilic
  than people with the opposite political lean?
\item
  In a particular school, is gender, race, income level, age, etc. the
  main driver of homophilic relationships?
\item
  In a citation network, professors in which disciplines are the most
  likely to cite papers written by professors in other disciplines?
\end{itemize}

  \clearpage 
\nocite{*}
\bibliographystyle{plainnat-updated} \bibliography{pioneer}

\end{document}